# Detection of Perineural Invasion in Prostate Needle Biopsies with Deep Neural Networks

*Peter Ström, M.Sc.[1], Kimmo Kartasalo, M.Sc.[1,2], Pekka Ruusuvuori, Ph.D.[2,3], Henrik Grönberg, M.D.[1,4], Hemamali Samaratunga, FRCPA[5], Brett Delahunt, M.D.[6], Toyonori Tsuzuki, M.D.[7], Lars Egevad, M.D.[8], and Martin Eklund, Ph.D.[1]*

Corresponding author: Dr. Martin Eklund; Department of Medical Epidemiology and Biostatistics, Karolinska Institutet, PO Box 281, SE-171 77 Stockholm, Sweden; martin.eklund@ki.se; +46 737121611

1. Department of Medical Epidemiology and Biostatistics, Karolinska Institutet, Stockholm, Sweden.
2. Faculty of Medicine and Health Technology, Tampere University, Tampere, Finland.
3. Institute of Biomedicine, University of Turku, Turku, Finland.
4. Department of Oncology, S:t Göran Hospital, Stockholm, Sweden.
5. Aquesta Uropathology and University of Queensland, Brisbane, QLD, Australia.
6. Department of Pathology and Molecular Medicine, Wellington School of Medicine and Health Sciences, University of Otago, Wellington, New Zealand.
7. Department of Surgical Pathology, School of Medicine, Aichi Medical University, Nagoya, Japan.
8. Department of Oncology and Pathology, Karolinska Institutet, Stockholm, Sweden.

## Abstract

**Background:** The detection of perineural invasion (PNI) by carcinoma in prostate biopsies has been shown to be associated with poor prognosis. The assessment and quantification of PNI is; however, labor intensive. In the study we aimed to develop an algorithm based on deep neural networks to aid pathologists in this task.
**Methods:** We collected, digitized and pixel-wise annotated the PNI findings in each of the approximately 80,000 biopsy cores from the 7,406 men who underwent biopsy in the prospective and diagnostic STHLM3 trial between 2012 and 2014. In total, 485 biopsy cores showed PNI. We also digitized more than 10% (n=8,318) of the PNI negative biopsy cores. Digitized biopsies from a random selection of 80% of the men were used to build deep neural networks, and the remaining 20% were used to evaluate the performance of the algorithm.
**Results:** For the detection of PNI in prostate biopsy cores the network had an estimated area under the receiver operating characteristics curve of 0.98 (95% CI 0.97-0.99) based on 106 PNI positive cores and 1,652 PNI negative cores in the independent test set. For the pre-specified operating point this translates to sensitivity of 0.87 and specificity of 0.97. The corresponding positive and negative predictive values were 0.67 and 0.99, respectively. For localizing the regions of PNI within a slide we estimated an average intersection over union of 0.50 (CI: 0.46-0.55).
**Conclusion:** We have developed an algorithm based on deep neural networks for detecting PNI in prostate biopsies with apparently acceptable diagnostic properties. These algorithms have the potential to aid pathologists in the day-to-day work by drastically reducing the number of biopsy cores that need to be assessed for PNI and by highlighting regions of diagnostic interest.

## Introduction

The identification of perineural invasion (PNI) by prostate carcinoma in prostate biopsies has been shown to be associated with poor outcomes.[1,2] In the United States each year approximately one million men undergo prostate biopsy and in various series PNI ranges from 7% to 33% of cases.[2,3] The intuitive reason for the poorer prognosis of men with PNI is that the cancer has invaded the perineural space of at least one nerve and it is via this route that tumor is able to spread outside the prostate. Despite increasing evidence of the prognostic significance of PNI, pathology reporting guidelines have so far not included PNI as a mandatory reporting element, although recently it has been included in prognostic guidelines for urologists[4,5]

A possible reason for PNI not being included in prostate pathology reporting guidelines may be that results from early studies regarding the prognostic significance of PNI were contradictory.[6] As a result it would seem a reasonable conclusion that the reporting of PNI could not be justified.[7] More recent studies have indicated that the identification of PNI is of prognostic significance and there is a growing body of evidence to suggest that PNI should be routinely reported in prostate biopsies.[6]

Workload issues are an on-going problem in pathology. Internationally the number of pathologists in clinical practice is in decline, while the breadth and complexity of pathology reporting is increasing. This is especially so in the case of prostate biopsies as the incidence of prostate cancer is increasing due to an increasing demand for informal/opportunistic screening for prostate cancer in an aging population. The workload issue is further compounded by the increasing number of biopsy cores that are taken as part of random sampling of the prostate for cancer. Similarly increasing numbers of cores are submitted from targeted biopsies that are designed to both diagnose and delineate the extent of malignancy. It is for this reason that recent initiatives have resulted in the development of artificial intelligence (AI) systems that have been designed to screen for cancer. The expectation here is that these will facilitate cancer diagnosis and assist pathologists in the routine screening of prostate cancer biopsies.[8]

Recently, two studies have demonstrated AI systems utilizing deep neural networks (DNN) to perform equivalently to expert uro-pathologists in grading prostate biopsies.[9,10] These networks have been trained through the screening of thousands of sections of both benign and malignant biopsies, including a variety of cancer morphologies. In these series cancers were diagnosed in foci of PNI; however, a drawback with these current networks is that they lack the capability to distinguish between PNI and invasive cancer. While PNI will be recognized as a malignant focus its true nature is overlooked, which means that potentially useful prognostic information is lost. In view of this it is apparent that if AI is to have a role in prostate cancer diagnosis there is a need for networks to be built that explicitly target relevant pathological features. Despite the obvious need to expand the role of AI systems in the diagnosis of prostate cancer, to the best of our knowledge, there are no studies that have examined the utility of using AI to target PNI.

This study is based upon a large series of prostate cancer biopsies accessioned prospectively as part of a trial to develop DNNs for the diagnosis of prostate cancer.[11,12] Specifically, we have utilized a subset of cases to develop and validate the automatic detection of PNI in prostate biopsies, with the aim of demonstrating clinically useful diagnostic properties.

# Methods

## Study design and participants

This study is based upon biopsy cores from men who participated in the STHLM3 trial (ISRCTN84445406). This was a prospective and population-based trial designed to evaluate a diagnostic model for prostate cancer. Patients in the trial were aged between 50 to 60 years and cases were accrued between May 28, 2012, and December 30, 2014. Formal diagnosis of prostate cancer was by 10-12 core transrectal ultrasound guided systematic biopsies.

## Slide preparation and digitization

The biopsy cores were formalin fixed, stained with haematoxylin and eosin and mounted on glass slides. Histologic assessment was undertaken by the study pathologist (LE) and pathological features including cancer grade and PNI were entered into a database. We then randomly selected 1,427 participants from which we retrieved 8,803 biopsy cores. The selection was stratified by grade to include a larger sample of high-grade cancers and all cases containing PNI (see Fig. 1 and Supplementary Appendix 1). From these, we randomly assigned 20% of the subjects to a test set to evaluate PNI.

## Slide annotations

Each slide containing PNI was re-assessed digitally by the study pathologist (LE) and all regions of PNI were annotated pixel-wise using QuPath.[13] In total there were 485 slides that contained at least one cancer with associated PNI. Binary masks of the slides were generated, and they acted as pixel-wise ground truth labels for training and validation purposes (see Supplementary Appendix 1).

## Deep neural networks

### Patch extraction

To train the DNNs on PNI morphology we extracted patches from each of the slides. Patch size was approximately 0.25 mm x 0.25 mm (see Supplementary Appendix 3), which was large enough to cover the size and shape of most of the nerves which showed infiltration by cancer. We evaluated different patch sizes within the training data with slightly lower performance (see Supplementary Appendix 4). To learn pixel-wise prediction we also extracted the corresponding region from the binary masks which acted as labels.

### Network architecture

Convolutional DNNs were used to classify patches (Xception) and to identify the regions in the biopsy where PNI was present (Unet).[14,15]

For classification, we used soft voting (i.e. averaging probabilities) from an ensemble of 10 networks to generate final patch wise probabilities. The highest probability for PNI among patches from a single core was used as a prediction score for classifying a core as PNI positive or negative at different operating points. Similarly, we used the highest predicted patch within a subject for subject-level classification.

For segmentation, i.e. pixel-wise prediction to identify the exact regions on each slide where the PNI was located, we first obtained pixel-wise predictions for each patch. We then re-mapped the predictions to the location from which each patch was extracted. We applied pixel-wise averaging across the slide for pixels with overlapping patches, and over all networks in the ensemble to generate probabilities for each pixel to contain PNI. Finally, we used an *a priori* specified threshold to classify each pixel in the slide as PNI positive or negative (see Supplementary Appendix 3).

### Statistical analysis

The receiver operating characteristic (ROC) curve and the area under the ROC curve (AUC) were used to evaluate the performance of the algorithm. In addition, we analyzed four pre-specified operating points (the index test and three alternative positivity criteria) on which we have reported sensitivity, specificity, positive predictive value (PPV), negative predictive value (NPV), and accuracy. For evaluating pixel-wise segmentation, we used intersection over union (IoU). Specifically, we used all predicted positive and true positive pixels for each core to measure the core wise IoU, and then reported the average of these IoUs across all PNI positive cores. All analyses (except IoU) were done both at individual core level and at subject level.

All confidence intervals (CIs) were two-sided with 95% confidence levels calculated from 1000 bootstrap samples. The networks and all analyses were implemented in Python (version 3.6.5) and TensorFlow (version 2.0.0).[16] For the Unet implementation we used the Python package segmentation_models with focal loss.[17]

## Results

In this study of 1,427 subjects 266 (18.6%) were positive for PNI. The PNI positive men generally had higher serum prostate specific antigen (PSA) levels prior to biopsy, were more likely to have positive findings on digital rectal examination and had cancers that were larger and of higher grade (see Table 1). From these subjects 8,803 slides were examined, of which 485 (5.5%) were positive for PNI.

The AUC for discriminating between PNI positive and negative was 0.98 (CI: 0.97-0.99) for individual slides (PNI positive = 106, PNI negative = 1,652), and 0.97 (CI: 0.93 - 0.99) for subjects (PNI positive = 52, PNI negative = 234), (Fig. 2). Sensitivity and specificity, positive and negative predictive values, and accuracy at the index test's operating point and three alternative operating points are shown in Table 2, both at the level of individual cores and at a subject level.

The estimated mean IoU across slides was 0.50 (CI: 0.46-0.55), which measures pixel-wise agreement between the study pathologist's annotation of PNI and the pixels classified as positive by the algorithm. For reference, Fig. 3 shows the individual PNI positive slide in the test set with IoU closest to the mean IoU.

# Discussion

Even though the propensity of prostate cancer to invade perineural spaces is well known, it is only comparatively recently that it has been shown to be independently associated with poor outcome. The presence of PNI appears to be of more prognostic significance when detected in biopsies and in accordance with this, urology practice guidelines support the reporting of PNI. Since the assessment of needle biopsies for the presence of PNI biopsies is tedious there is well recognized inter-observer error, which itself may have contributed to the confusion relating to the prognostic significance of PNI. Currently, pathology reporting guidelines, issued by both international bodies and jurisdictional pathology authorities, do not include PNI as a required element.[18–20] In view of the increasing evidence of the utility of PNI detection as a prognostic parameter this is likely to change.[21] In some cases identification of PNI may be problematic, with foci of infiltration of the perineural space being difficult to distinguish from stromal bundles or collagenous micronodules situated adjacent to malignant glands. In this context DNNs may assist in the detection of PNI as they appear to reach a more consistent level of performance when compared to the subjective observations of diagnostic pathologists. Since DNNs are consistent in their decisions and are easily distributed, they also have potential value as a teaching tool.

In this study we have demonstrated a novel deep learning system which we have shown to detect PNI in prostate cancer biopsies with high AUC. Use of the system may also assist pathologists by suggesting regions of interest for PNI in a biopsy slide. We believe that with the addition of further cases the diagnostic specificity of the reporting system can be further improved. Specifically, additional cases would permit further calibration of the system to more accurately handle difficult cases based on the diagnostic input from expert pathologists.

The main strength of this study is that we have identified, digitized and annotated all foci of PNI in each of more than 80,000 biopsy cores from all men who underwent biopsy as part of the STHLM3 trial. Since this trial was based upon a randomized population-based selection of men, there is a strong probability that the tumors sampled have displayed a broad spectrum of morphologies. This in turn suggests that the results may be generalized, and the diagnostic algorithms developed in this study are applicable to other populations. Further strengths of the study are that our reporting network was validated by a large independent test set and that the pathology of all the cases in the series was evaluated and reported by a single specialist prostate pathologist.

The main limitation of the study is that, due to the high cost of digitizing the samples, we could not include biopsies from all the subjects that participated in the STHLM3 trial in this PNI study. We felt that this was untenable due to the relatively low value that would have been added to the study by including all the numerous cores with no tumor and those with low grade cancers.

The selection of cases for study was random but cases were stratified by grade to include sufficient high-grade cancers in the series. This was considered necessary as high-grade prostate cancers are only infrequently encountered in screening trials. This makes it difficult to interpret the predictive values, which depend on the prevalence of positive and negative cases. For example, if we had not over-sampled positive cases the already high NPV would likely be even higher and the somewhat low PPV would likely be even lower. Even if the PPV appears low – as approximately half of the predicted positive slides do not contain PNI – it should be noted that in the series PNI was not present in most cases. Despite this, only assessing the slides which were predicted as positive would result in a substantial reduction in work for the reporting pathologist. Importantly, the AUCs (sensitivity and specificity) were not confounded by artificial oversampling of positive cases.

Another limitation in the study was a difficulty in the interpretation of IoU. In this study we chose to define intersection and union as all pixels of PNI in a slide as a single object. This did not consider relative sizes of PNI within a specific slide. For example, given a slide containing a large and a small PNI focus, one would obtain a higher IoU by fully detecting the large focus and fully failing to identify the small focus, rather than by partially detecting both foci. We would argue that the latter would still be more desirable when using the system as a diagnostic aid, but this is not captured by the IoU metric. Finally, we have not tested the algorithm on external data. We know from other studies involving whole slide images that one can expect some loss in performance on an external test set, likely due to differences in laboratory staining protocols, subjectivity in the assessment of pathology and potentially the use of different types of scanners to digitize the biopsy cores. To overcome these limitations, it is preferable to include a large sample in training the networks, specimens from different laboratories, a variety of types of image scanners and a wide range of prostate tumors showing differing morphologies.

Neural networks have shown excellent results in the automation of the grading of cancers in prostate biopsies. An important limitation; however, is an inherent lack of flexibility. Networks only perform the task that they are trained on and do not reveal information relating to other findings. Moving towards a fully automated assessment of biopsies, we need networks that can address the interpretation of features additional to grading. All potential tasks do not need to be incorporated in a single network but can be implemented through several separate networks with each performing a single task. The evolving digital revolution of pathology will undoubtedly give rise to abundant data that can be utilized for training such specific networks.

## Conclusion

This study has demonstrated that deep neural networks can screen appropriately for perineural invasion by cancer in prostate biopsies. This has the potential to reduce the workload for pathologists. Application of such systems will also allow for automatic interpretation of large datasets, which can be utilized to increase knowledge as to the reasons for the relationship between perineural invasion by prostate cancer and poor patient outcome.

# References

1. Ström P, Nordström T, Delahunt B, et al. Prognostic value of perineural invasion in prostate needle biopsies: a population-based study of patients treated by radical prostatectomy. J Clin Pathol 2020;
2. Wu S, Lin X, Lin SX, et al. Impact of biopsy perineural invasion on the outcomes of patients who underwent radical prostatectomy: a systematic review and meta-analysis. Scand J Urol 2019;53(5):287–94.
3. Loeb S, Carter HB, Berndt SI, Ricker W, Schaeffer EM. Complications after prostate biopsy: Data from SEER-Medicare. J Urol 2011;
4. Grignon DJ. Prostate cancer reporting and staging: needle biopsy and radical prostatectomy specimens. Mod Pathol 2018;
5. Mottet N, Bellmunt J, Bolla M, et al. EAU-ESTRO-SIOG Guidelines on Prostate Cancer. Part 1: Screening, Diagnosis, and Local Treatment with Curative Intent. Eur Urol 2017;
6. Delahunt B, Murray JD, Steigler A, et al. Perineural invasion by prostate adenocarcinoma in needle biopsies predicts bone metastasis: Ten year data from the TROG 03.04 RADAR Trial. Histopathology 2020;(in press).
7. Metter DM, Colgan TJ, Leung ST, Timmons CF, Park JY. Trends in the US and Canadian Pathologist Workforces From 2007 to 2017. JAMA Netw Open [Internet] 2019;2(5):e194337. Available from: http://dx.doi.org/10.1001/jamanetworkopen.2019.4337
8. Egevad L, Ström P, Kartasalo K, et al. The Utility of Artificial Intelligence in the Assessment of Prostate Pathology. Histopathology 2020;In press.
9. Strom P, Kartasalo K, Olsson H, et al. Artificial intelligence for diagnosis and grading of prostate cancer in biopsies: a population-based, diagnostic study. Lancet Oncol 2020;
10. Bulten W, Pinckaers H, van Boven H, et al. Automated deep-learning system for Gleason grading of prostate cancer using biopsies: a diagnostic study. Lancet Oncol [Internet] 2020;21(2):233–41. Available from: http://dx.doi.org/10.1016/S1470-2045(19)30739-9
11. Grönberg H, Adolfsson J, Aly M, et al. Prostate cancer screening in men aged 50-69 years (STHLM3): A prospective population-based diagnostic study. Lancet Oncol [Internet] 2015;16(16):1667–76. Available from: http://dx.doi.org/10.1016/S1470-2045(15)00361-7
12. Ström P, Nordström T, Aly M, Egevad L, Grönberg H, Eklund M. The Stockholm-3 Model for Prostate Cancer Detection: Algorithm Update, Biomarker Contribution, and Reflex Test Potential. Eur Urol [Internet] 2018;74(2):204–10. Available from: http://dx.doi.org/10.1016/j.eururo.2017.12.028
13. Bankhead P, Loughrey MB, Fernández JA, et al. QuPath: Open source software for digital pathology image analysis. Sci Rep [Internet] 2017;7(1):16878. Available from: http://dx.doi.org/10.1038/s41598-017-17204-5
14. Chollet F. Xception: Deep Learning with Depthwise Separable Convolutions [Internet]. 2017 IEEE Conf. Comput. Vis. Pattern Recognit. 2017;Available from: http://dx.doi.org/10.1109/cvpr.2017.195 LB  - Maqo
15. Ronneberger O, Fischer P, Brox T. U-Net: Convolutional Networks for Biomedical Image Segmentation [Internet]. Lect. Notes Comput. Sci. 2015;234–41. Available from: http://dx.doi.org/10.1007/978-3-319-24574-4_28 LB  - XxKs
16. Martín Abadi, Ashish Agarwal, Paul Barham, et al. TensorFlow: Large-Scale Machine

Learning on Heterogeneous Systems [Internet]. 2015;Available from: https://www.tensorflow.org/
17. Lin T-Y, Goyal P, Girshick R, He K, Dollar P. Focal Loss for Dense Object Detection. IEEE Trans Pattern Anal Mach Intell [Internet] 2020;42(2):318–27. Available from: http://dx.doi.org/10.1109/TPAMI.2018.2858826
18. International Collaboration on Cancer Reporting. Prostate cancer histopathology reporting guide – radical prostatectomy specimen. 2018. http://www.iccr-cancer.org/datasets/published-datasets/urinary-male-genital. :Accessed: 2020-03-30.
19. College of American Pathologists. Cancer protocol templates. 2018. http://www.cap.org/web/oracle/webcenter/portalapp/pagehierarchy/cancer_protocol_templates.jspx?_adf.ctrlstate=7ld3v1dp4_14&_afrLoop=100775365203342#! :Accessed: 2020-03-30.
20. Royal College of Pathologists of Australasia. Cancer protocols. 2018. www.rcpa.edu.au/Library/Practising-Pathology/Structured-Pathology-Reporting-of-Cancer/Cancer-Protocols. :Accessed: 2020-03-30.
21. Egevad L, Judge M, Delahunt B, et al. Dataset for the reporting of prostate carcinoma in core needle biopsy and transurethral resection and enucleation specimens: recommendations from the International Collaboration on Cancer Reporting (ICCR). Pathology 2019;51(1):11–20.

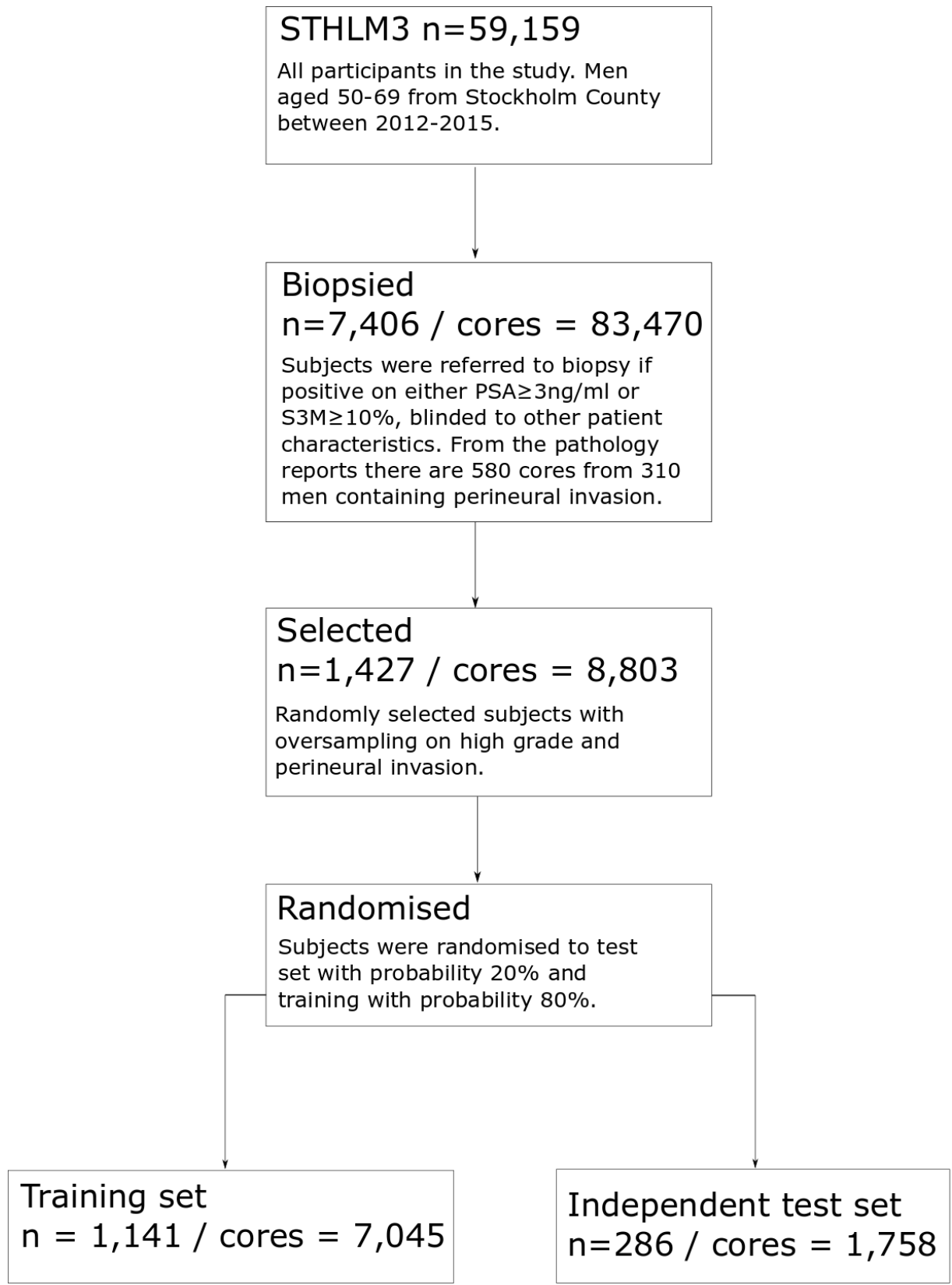

**Figure 1**: Patient flow diagram.

|  | PNI positive men (n=266) No. (%) | PNI negative men (n=1161) No. (%) |
| --- | --- | --- |
| **Age** | | |
| <49 yr | 2 (0.8) | 4 (0.3) |
| 50-54 yr | 13 (5.0) | 89 (7.7) |
| 55-59 yr | 35 (13.6) | 170 (14.7) |
| 60-64 yr | 83 (32.2) | 303 (26.2) |
| 65-69 yr | 119 (46.1) | 564 (48.7) |
| ≥70 yr | 6 (2.3) | 28 (2.4) |
| **PSA** | | |
| <3 ng/mL | 46 (17.8) | 271 (23.4) |
| 3-5 ng/mL | 78 (30.2) | 531 (45.9) |
| 5-10 ng/mL | 71 (27.5) | 261 (22.5) |
| ≥10 ng/mL | 63 (24.4) | 95 (8.2) |
| **Digital rectal examination** | | |
| Abnormal | 109 (42.2) | 119 (10.3) |
| Normal | 149 (57.8) | 1039 (89.7) |
| **Prostate volume** | | |
| <35 mL | 138 (53.5) | 487 (42.1) |
| 35-50 mL | 74 (28.7) | 382 (33.0) |
| ≥50 mL | 46 (17.8) | 289 (25.0) |
| **Cancer length** | | |
| No cancer | 0 (0.0) | 176 (15.2) |
| 0-1 mm | 4 (1.6) | 169 (14.6) |
| 1-5 mm | 13 (5.0) | 335 (28.9) |
| 5-10 mm | 37 (14.3) | 170 (14.7) |
| >10 mm | 204 (79.1) | 308 (26.6) |
| **ISUP grade†** | | |
| Benign | 0 (0.0) | 176 (15.2) |
| ISUP 1 (3 + 3) | 40 (15.5) | 522 (45.1) |
| ISUP 2 (3 + 4) | 96 (37.2) | 248 (21.4) |
| ISUP 3 (4 + 3) | 58 (22.5) | 89 (7.7) |
| ISUP 4 (4 + 4, 3 + 5, 5 + 3) | 25 (9.7) | 68 (5.9) |
| ISUP 5 (4 + 5, 5 + 4, 5 + 5) | 39 (15.1) | 55 (4.7) |

|  | PNI positive slides (n=485) No. (%) | PNI negative slides (n=8318) No. (%) |
| --- | --- | --- |
| **Cancer length** | | |
| No cancer | 0 (0.0) | 4712 (56.6) |
| 0-1 mm | 69 (14.2) | 1121 (13.5) |
| 1-5 mm | 134 (27.6) | 1547 (18.6) |
| 5-10 mm | 170 (35.1) | 698 (8.4) |
| >10 mm | 112 (23.1) | 240 (2.9) |
| **ISUP grade†** | | |
| Benign | 0 (0.0) | 4712 (56.6) |
| ISUP 1 (3 + 3) | 97 (20.0) | 1892 (22.7) |
| ISUP 2 (3 + 4) | 130 (26.8) | 680 (8.2) |
| ISUP 3 (4 + 3) | 81 (16.7) | 321 (3.9) |
| ISUP 4 (4 + 4, 3 + 5, 5 + 3) | 85 (17.5) | 469 (5.6) |
| ISUP 5 (4 + 5, 5 + 4, 5 + 5) | 92 (19.0) | 244 (2.9) |

**Table 1**: **(top)** Patient characteristics. There were 11 patients (8 of 266 PNI positive men) on whom we could not retrieve clinical information. **(bottom)** Slide characteristics. There was no missing information. † The values in parentheses are the Gleason scores.

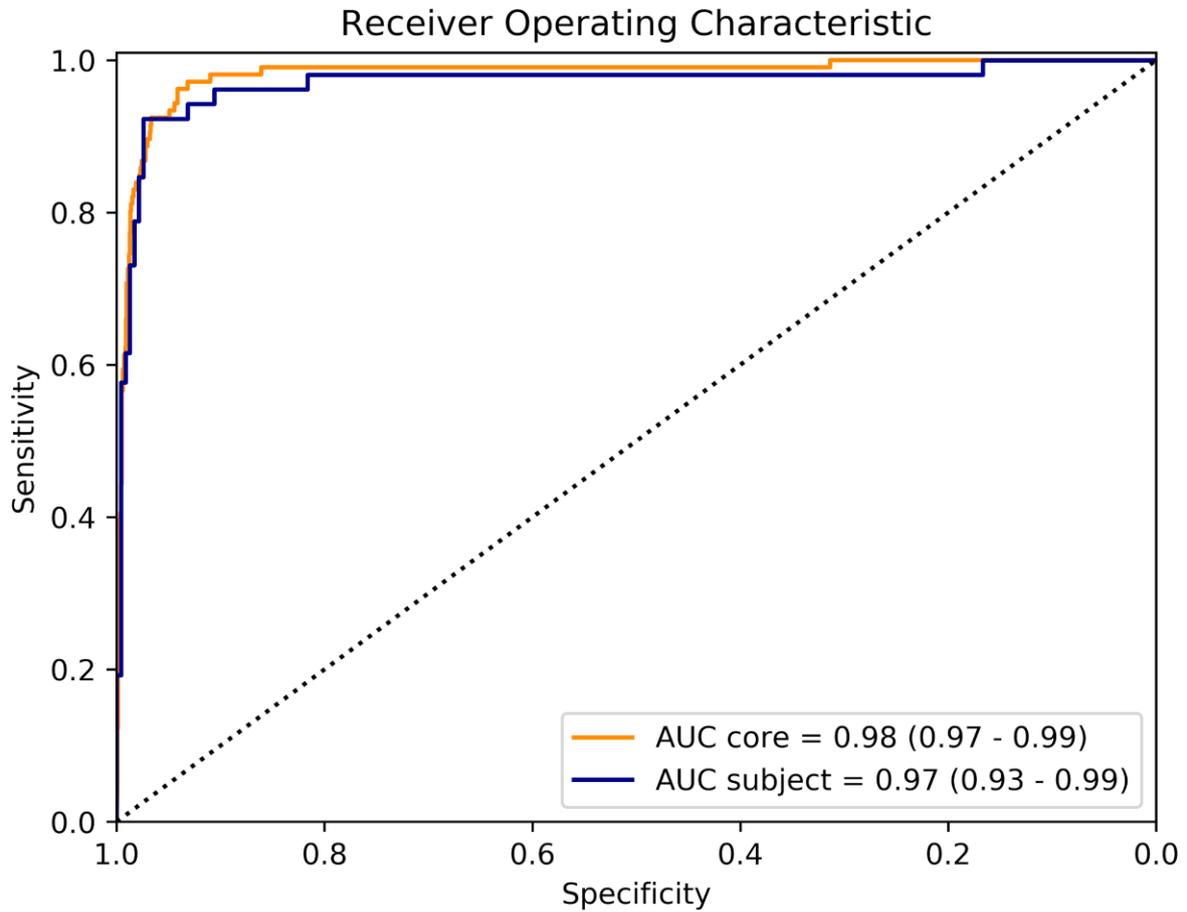

**Figure 2**: Performance of the network to discriminate between PNI and non-PNI in individual cores (orange) and in subjects (blue). The curves are based on n=1,758 (n=106 positive) cores and n=286 (n=52 positive) subjects. The values in parentheses are confidence intervals.

|  | Operating point | Sensitivity | Specificity | PPV | NPV | Accuracy |
|---|---|---|---|---|---|---|
| **Cores** | 0.99 | 0.82 | 0.98 | 0.78 | 0.99 | 0.97 |
| Postive n = 106 | 0.95 (index test) | 0.87 | 0.97 | 0.67 | 0.99 | 0.97 |
| Negative n = 1652 | 0.90 | 0.92 | 0.96 | 0.60 | 0.99 | 0.96 |
|  | 0.85 | 0.92 | 0.95 | 0.54 | 0.99 | 0.95 |
|  |  |  |  |  |  |  |
| **Subjects** | 0.99 | 0.92 | 0.96 | 0.84 | 0.98 | 0.95 |
| Postive n = 52 | 0.95 (index test) | 0.94 | 0.91 | 0.69 | 0.99 | 0.91 |
| Negative n = 234 | 0.90 | 0.96 | 0.85 | 0.60 | 0.99 | 0.87 |
|  | 0.85 | 0.96 | 0.84 | 0.57 | 0.99 | 0.86 |

**Table 2**: Diagnostic metrics for the network. The operating points are alternative thresholds for positivity. The point marked (index test) is the value on which the algorithm is intended to be used, and the other three show the diagnostic properties of the model if a different sensitivity to specificity relationship is preferred. PPV = Positive Predictive Value, NPV = Negative Predictive Value.

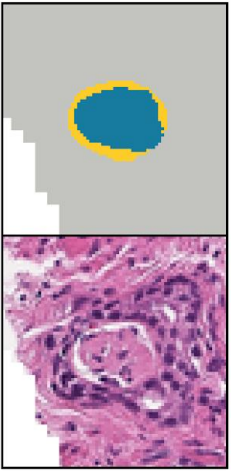
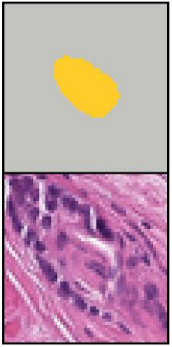
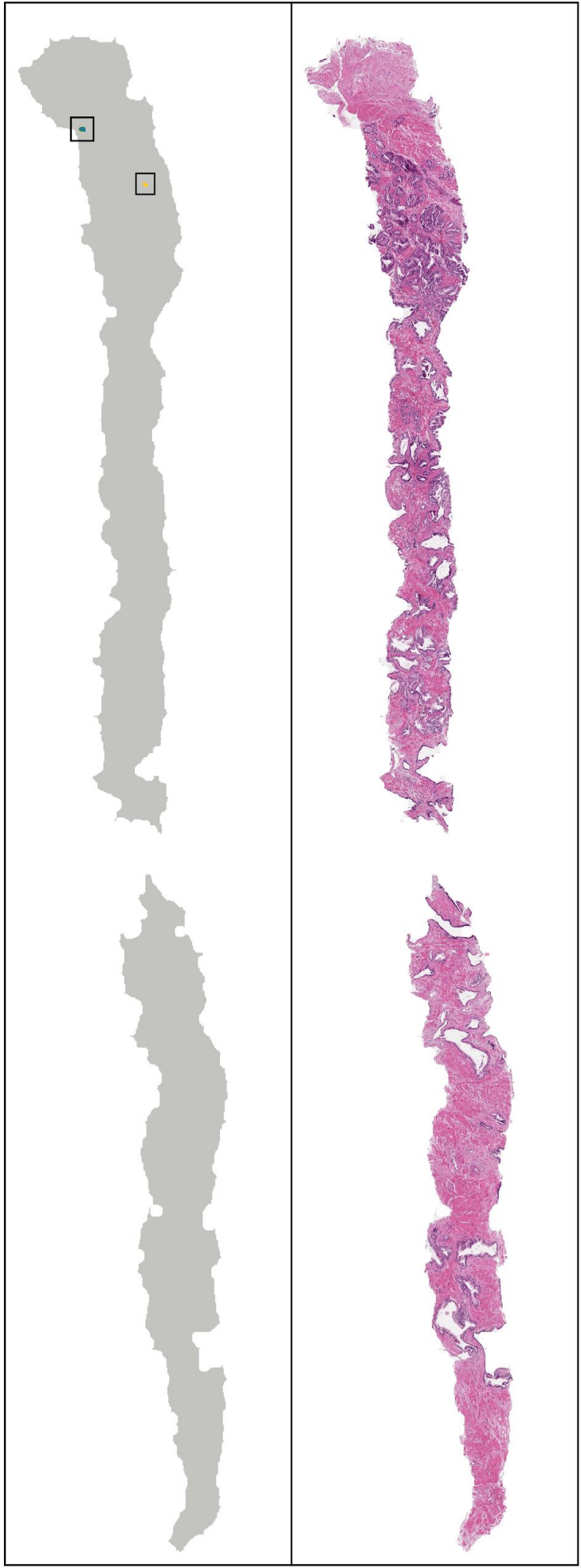

**Figure 3**: Illustration of PNI segmentation on the biopsy core with IoU (0.51) closest to the overall mean IoU (0.50) reported. The H&E stained biopsy **(right)** and the corresponding predicted pixel-wise classification and ground truth **(middle)**, and two highlighted regions **(left)** are shown. The regions both annotated by the pathologist and classified as positive by the network (i.e. the intersection) is colored blue, and the regions not annotated but still classified positive by the network are yellow. In this example there were no regions annotated but not classified as positive. All pixels positive by either the pathologist or the network form the union (i.e. the denominator in the IoU).

# Supplementary

*Peter Ström, M.Sc.[1], Kimmo Kartasalo, M.Sc.[1,2], Pekka Ruusuvuori, Ph.D.[2,3], Henrik Grönberg, M.D.[1,4], Hemamali Samaratunga,* FRCPA[5], *Brett Delahunt, M.D.[6], Toyonori Tsuzuki, M.D.[7], Lars Egevad, M.D.[8], and Martin Eklund, Ph.D.[1]*

Corresponding author: Dr. Martin Eklund; Department of Medical Epidemiology and Biostatistics, Karolinska Institutet, PO Box 281, SE-171 77 Stockholm, Sweden; martin.eklund@ki.se; +46 737121611

1. Department of Medical Epidemiology and Biostatistics, Karolinska Institutet, Stockholm, Sweden.
2. Faculty of Medicine and Health Technology, Tampere University, Tampere, Finland.
3. Institute of Biomedicine, University of Turku, Turku, Finland.
4. Department of Oncology, S:t Göran Hospital, Stockholm, Sweden.
5. Aquesta Uropathology and University of Queensland, Brisbane, QLD, Australia.
6. Department of Pathology and Molecular Medicine, Wellington School of Medicine and Health Sciences, University of Otago, Wellington, New Zealand.
7. Department of Surgical Pathology, School of Medicine, Aichi Medical University, Nagoya, Japan.
8. Department of Oncology and Pathology, Karolinska Institutet, Stockholm, Sweden.

# Appendix 1: Data acquisition

We used the STHLM3 register to select subjects whose biopsies we digitized. The first 500 men who were diagnosed in the STHLM3 trial were selected. This contributed the majority of slides, n=5,662. Since it was a screening-by-invitation trial, there were relatively few high-grade cases in this selection. To increase the coverage of the more rare high-grade cases and their associated morphologies, we included all subjects (their positive cores and a randomly selected negative core) with at least one core graded as Gleason Score (GS) 4+4 or 5+5, and a random selection of 497 subjects with at least one core graded 3+3. To increase the sample diversity of benign prostate tissue, we included 139 randomly selected cancer-free subjects from which we included one randomly selected core each. At this stage we had selected 1,289 participants from which we retrieved 8,571 biopsy cores. Finally, we included all cores in the STHLM3 register which had PNI reported which concluded a total of 8,803 slides from 1,427 subjects. In some of these, the study pathologist did not manage to reproduce the PNI finding. This could be due to difficult to judge cases or the difference of using microscopy (the original assessment) and digital pathology (the annotation of PNI for this study). These slides were included as negative cases.

The study pathologist (L.E.) subsequently highlighted each lesion of PNI pixel-wise in the digitized cores that were reported to contain PNI. This was done in the open source pathology software QuPath.[1] From these annotations we created binary masks, i.e. images indicating the presence or absence of PNI for each pixel of the associated whole slide images. The binary masks acted as pixel-wise ground truth labels when training and evaluating the networks.

There were two types of scanners used for digitizing the slides: Hamamatsu C9600-12 scanner running NDP.scan software v. 2.5.86 (Hamamatsu Photonics, Hamamatsu, Japan) and Aperio ScanScope AT2 scanner running Aperio Image Library v. 12.0.15 software (Leica Biosystems, Wetzlar, Germany). At full-resolution (20X) the pixel sizes were 0.45202 μm (Hamamatsu) and 0.5032 μm (Aperio). The RGB images were stored at 8-bits per channel in NDPI (Hamamatsu) and SVS (Aperio) format.

# Appendix 2: Pre-processing

The scanned images typically contain two consecutive sections from a single biopsy core. The annotation of PNI was done on one of these two sections arbitrarily chosen by the study pathologist. To only include the annotated section in the training, we semi-automatically removed the unannotated one. In the cases where the annotated section was located in its entirety within one half of the image, and the unannotated section within the other half, we retained the image half containing the annotated section. Otherwise, removal of the unannotated section was performed manually.

Images were downsampled by a factor of 16 and converted from RGB to grayscale in accordance with the NTSC standard by calculating the weighted sum 0.2989 x R + 0.5870 x G + 0.1140 x B for each pixel. The tissue was segmented using Laplacian filtering, followed by thresholding the absolute magnitude of the resulting response using Otsu's method.[2] This resulted in binary masks indicating the tissue regions. For further details see Supplementary of Ström et al.[3]

To create label masks, that is, masks with values 0 for background, 1 for non-PNI tissue, and 2 for PNI, we exported binary masks from QuPath for each unique region annotated by the pathologist. We also stored the pixel coordinates from which they were extracted. Finally, we removed the complementary section in the mask as described above.

# Appendix 3: Neural Networks

## Patch extraction

For classification, we cropped patches with dimensions of 598 x 598 pixels at the highest resolution (20X) of the whole slide images, corresponding to roughly 300 μm x 300 μm. The patches were then labeled as PNI positive if they contained at least one pixel of PNI based on the binary masks.

For segmentation, we only used the slides positive for PNI and cropped patches of size 512 x 512 pixels at 20X from them, corresponding to 250 μm x 250 μm. The reason for using a different patch size than for classification was to match the different network architectures' default input sizes by factors of 2. In addition to cropping the tissue, we also cropped the corresponding region from the binary mask. These patches of the masks acted as the target variable when training the model for pixel-wise segmentation of PNI.

Since most parts of the image are background, we only considered patches with at least half of the pixels containing tissue according to the tissue mask. Patches were systematically cropped from all parts of the whole slide images containing tissue, with 50% overlap between adjacent patches. Finally, the patches were stored on disk as .jpg using JPEG compression with 80% quality.

## Classification network

The deep neural network (DNN) used for classification of PNI on the patch level was Xception.[4] Ten models were trained in an identical way (except for randomization in initiating the model weights and the sampling and sample order of patches), and an average of these models' predictions was used as the final patch level prediction. For slide-level prediction, the maximum over the predictions of all patches from a slide was taken to represent the entire slide. Similarly, slide-level predictions were aggregated into subject-level predictions by taking the maximum over all slides from a given subject.

For training the DNNs we used a batch size of 8, randomly initiated weights, partial class balancing via oversampling of the rare positive class (2 benign patches sampled per one positive patch) to counter the extreme class imbalance of the data, binary crossentropy loss, Adam as optimizer with learning rate 0.001 and with other parameters set to default values, and trained for 100 epochs.[5] We augmented the data with random horizontal flips and random rotations of 90, 180, and 270 degrees.

## Segmentation network

For segmentation we applied Unet on the patches.[6] Similarly to the classification models, we used an ensemble of 10 DNNs. The prediction patches (i.e. pixel-wise predictions for PNI for an image patch) were mapped to their original positions in the image to produce a prediction image corresponding to each whole slide image. A threshold was used to generate a binary version of the prediction image.

For training the DNNs we used a batch size of 8, randomly initiated weights, partial class balancing via oversampling of the rare positive class (4 benign patches sampled per one positive patch) to counter the extreme class imbalance of the data, Adam as optimizer with a learning rate of 0.0001 and other parameters set to default values, focal binary loss and trained for 20 epochs.[7] We augmented the data with random horizontal flips and random rotations of 90, 180, and 270 degrees. Pixel-wise probabilities were averaged across the models in the ensemble and scaled from a floating-point range of [0, 1] to an unsigned 8-bit integer range of [0, 255]) to store the information as heatmap images. Finally, a pre-specified threshold of 75 was used to classify each pixel as positive. The reason that this was lower than a corresponding probability of 0.5 was to err on the side of sensitivity rather than specificity, since we argue that it is better to highlight additional potential foci rather than too few in the case of pixel-wise visualization.

Hardware and software

Computations were performed on a graphics processing unit (GPU) cluster (Tampere Center for Scientific Computing, Tampere, Finland) equipped with 28 x Tesla V100 and 32 x Tesla P100 GPUs (Nvidia, Santa Clara, CA, USA) distributed on 15 nodes. The GPUs were running Nvidia driver 440.64, CUDA 10.0.130 and cuDNN 7.6.0. The nodes were equipped with either a 20-core Xeon E5-2640 v4 or a 24-core Xeon Silver 4116 CPU (Intel, Santa Clara, CA, USA) and 254 GB, 385 GB or 785 GB of RAM. Each node was equipped with a local SSD disk. We used 32-bit floating point precision for GPU computation.

We used OpenSlide (v. 3.4.1) via the Python interface (v. 1.1.1) to access the images.[8] MATLAB R2017b (The MathWorks, Natick, MA, USA) and python were used to create the necessary masks for pixel-wise labeling of tissue and PNI. DNNs were implemented in Python 3.6.4 and TensorFlow (v 2.0.0).[9]

# Appendix 4: Complementary results

We evaluated a number of design choices on training data. The evaluation was performed on a fixed validation split of 20% of the training data, using the remaining data for training. The split was performed on subject level.

For classification we compared several DNN architectures: Inception V3, Xception, ResNet, InceptionResNetV2, NASNet, EfficentNet (B7)[10–15] Of these, Inception V3 and Xception generally had the best performance across various hyperparameter settings and appeared most reliable in terms of few severe drops in performance (data not shown). Xception exhibited slightly better performance than Inception V3. A comparison between the two architectures is shown in Figure S1.

Moreover, we evaluated several patch sizes and resolution combinations (see three well performing combinations in Figure S1). In addition, we evaluated different strides for extracting patches and observed that a stride of 299 pixels (that is, 50% overlap between patches of 598 x 598 pixels), was superior to a stride of 150 pixels (data not shown).

For segmentation, we compared several backbones for Unet: VGG16, ResNet, ResNeXt, Inceptionv3, and EfficientNet (B7).[16] Inception V3 was chosen based on consistent high performance on various sampling strategies and learning rates (data not shown). We also compared several loss functions (see Figure S2 and S3).

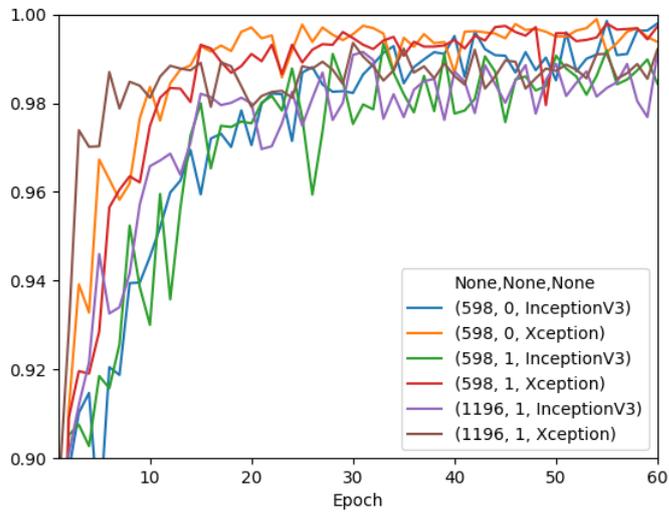

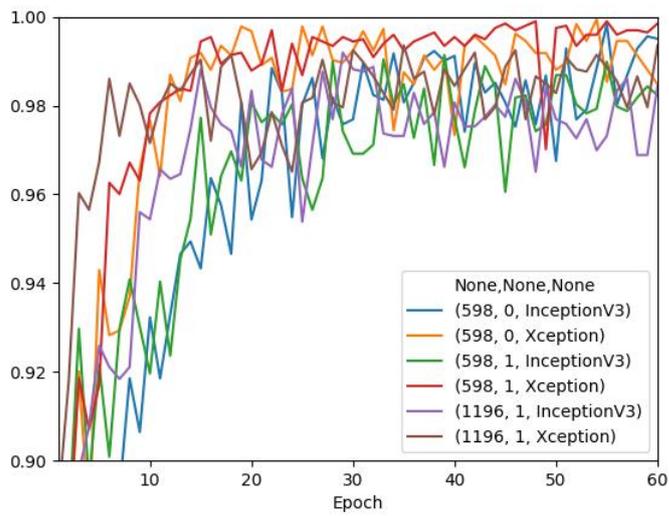

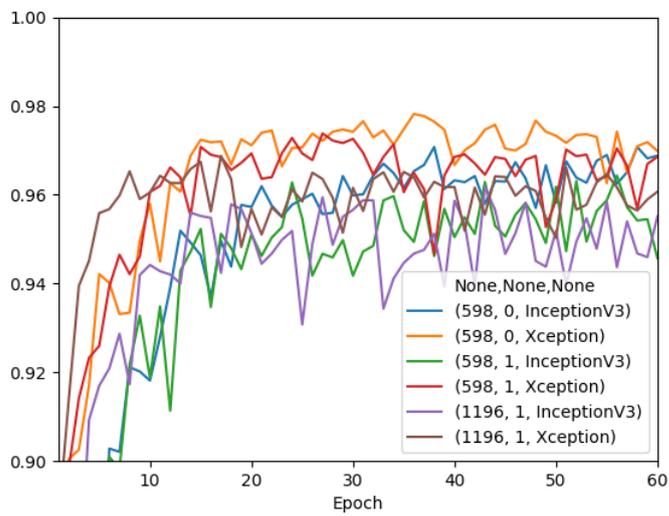

**Figure S1**: AUC for PNI classification using different patch sizes (598 and 1196 pixels), resolution (20X and 10X), and architecture (Inception V3 and Xception). From top: slide level, subject level, and patch level.

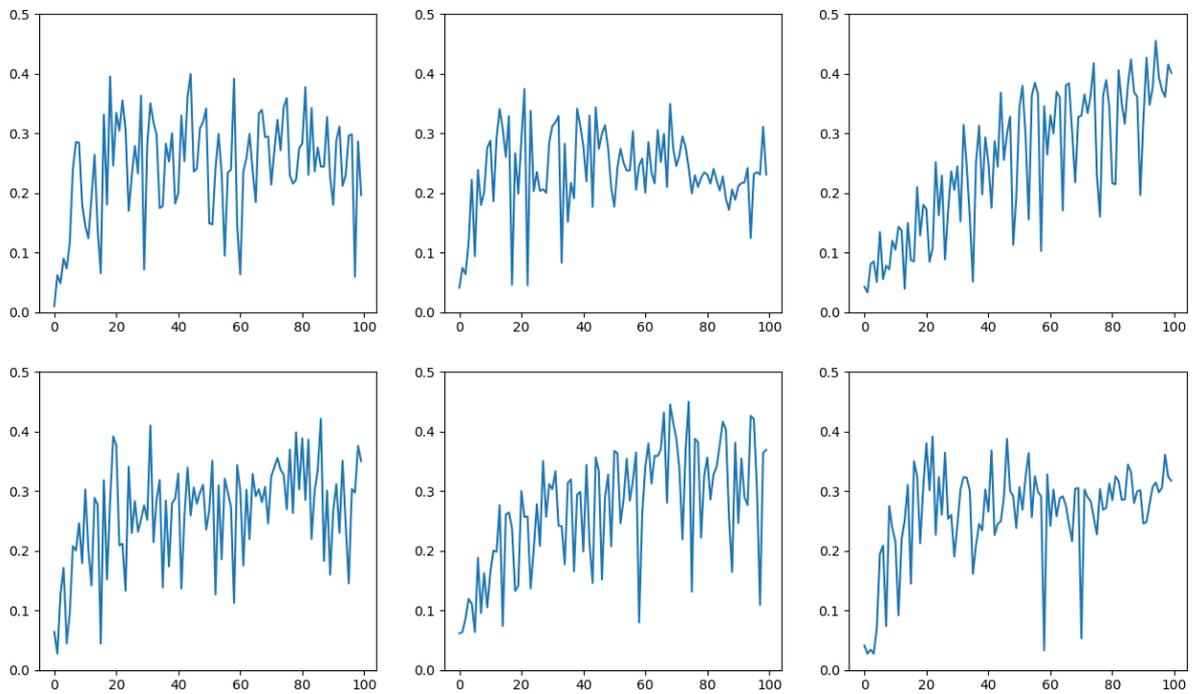

**Figure S2:** Intersection over union on patch level for various loss functions. **(First row; from left)** binary focal Jaccard loss, binary crossentropy, and Jaccard loss. **(Second row; from left)** binary focal dice loss, dice loss, and binary focal loss. The x-axis is number of epochs.

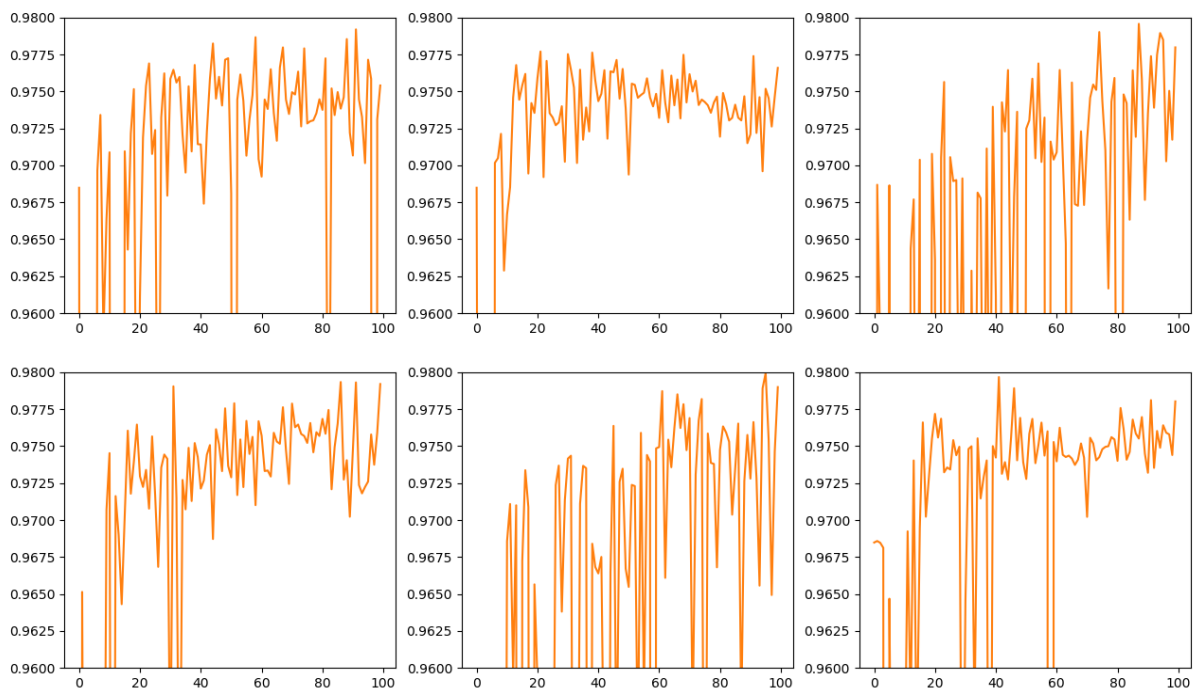

**Figure S3:** Accuracy on patch level for various loss functions. **(First row; from left)** binary focal Jaccard loss, binary crossentropy, and Jaccard loss. **(Second row; from left)** binary focal dice loss, dice loss, and binary focal loss. The x-axis is number of epochs.

# References

1. Bankhead, P. *et al.* QuPath: Open source software for digital pathology image analysis. *Sci. Rep.* **7**, 16878 (2017).
2. Otsu, N. A Threshold Selection Method from Gray-Level Histograms. *IEEE Trans. Syst. Man. Cybern.* **9**, 62–66 (1979).
3. Strom, P. *et al.* Artificial intelligence for diagnosis and grading of prostate cancer in biopsies: a population-based, diagnostic study. *Lancet. Oncol.* (2020). doi:10.1016/S1470-2045(19)30738-7
4. Chollet, F. Xception: Deep Learning with Depthwise Separable Convolutions. *2017 IEEE Conference on Computer Vision and Pattern Recognition (CVPR)* (2017). doi:10.1109/cvpr.2017.195
5. Kingma, D. P. & Ba, J. Adam: a method for stochastic optimization. CoRR abs/1412.6980 (2014). (2014).
6. Ronneberger, O., Fischer, P. & Brox, T. U-Net: Convolutional Networks for Biomedical Image Segmentation. *Lecture Notes in Computer Science* 234–241 (2015). doi:10.1007/978-3-319-24574-4_28
7. Lin, T.-Y., Goyal, P., Girshick, R., He, K. & Dollar, P. Focal Loss for Dense Object Detection. *IEEE Trans. Pattern Anal. Mach. Intell.* **42**, 318–327 (2020).
8. Goode, A., Gilbert, B., Harkes, J., Jukic, D. & others. OpenSlide: A vendor-neutral software foundation for digital pathology. *J. Pathol.* (2013).
9. Martín Abadi *et al.* TensorFlow: Large-Scale Machine Learning on Heterogeneous Systems. (2015).
10. Szegedy, C., Vanhoucke, V., Ioffe, S., Shlens, J. & Wojna, Z. Rethinking the Inception Architecture for Computer Vision. *Proceedings of the IEEE Computer Society Conference on Computer Vision and Pattern Recognition* **2016**-**Decem**, 2818–2826 (2016).
11. Chollet, F. Xception: Deep Learning with Depthwise Separable Convolutions. *2017 IEEE Conference on Computer Vision and Pattern Recognition (CVPR)* (2017). doi:10.1109/cvpr.2017.195
12. He, K., Zhang, X., Ren, S. & Sun, J. *Deep Residual Learning for Image Recognition*. (2016).
13. Szegedy, C., Ioffe, S., Vanhoucke, V. & Alemi, A. A. *Inception-v4, Inception-ResNet and the Impact of Residual Connections on Learning. Thirty-First AAAI Conference on Artificial Intelligence* (2017).
14. Zoph, B., Brain, G., Vasudevan, V., Shlens, J. & Le Google Brain, Q. V. *Learning Transferable Architectures for Scalable Image Recognition. openaccess.thecvf.com*
15. Tan, M. & Le, Q. V. EfficientNet: Rethinking model scaling for convolutional neural networks. in *36th International Conference on Machine Learning, ICML 2019* **2019**-**June**, 10691–10700 (International Machine Learning Society (IMLS), 2019).
16. Simonyan, K. & Zisserman, A. Very deep convolutional networks for large-scale image recognition. in *3rd International Conference on Learning Representations, ICLR 2015 - Conference Track Proceedings* (International Conference on Learning Representations, ICLR, 2015).